\documentclass{article}
\usepackage{arxiv}
\usepackage{graphics}
\usepackage{booktabs}       
\usepackage{amsfonts}       
\usepackage{nicefrac}       
\usepackage{microtype}      
\usepackage{lipsum}		
\usepackage{tabularray}
\usepackage{outlines}
\usepackage{graphicx}
\usepackage{natbib}
\usepackage{doi}
\usepackage{amsmath}
\usepackage{comment}
\usepackage{authblk}
\usepackage{subfig}

\usepackage{comment}
\usepackage{xcolor}
\usepackage{alltt}

\usepackage{graphicx,wrapfig,lipsum}
\usepackage{setspace}
\definecolor{indigo}{HTML}{4B69C6}
\definecolor{maroon}{HTML}{AA3732}
\definecolor{dgreen}{HTML}{319332}
\definecolor{tpurple}{HTML}{7A3E9D}
\definecolor{iibrown}{HTML}{9C5D27}
\definecolor{igrey}{HTML}{D3D3D4}

  \DeclareFontEncoding{FML}{}{}
  \DeclareFontSubstitution{FML}{fncmi}{m}{it}
  \DeclareSymbolFont{fourierletters}{FML}{fncmi}{m}{it}
  \DeclareMathSymbol{\freq}{\mathalpha}{fourierletters}{`f}

\begin{document}

\title{Formal verification of isothermal chemical reactors}

\author[1]{Parivash Feyzishendi}
\author [2]{Sophia Hamer}
\author[1]{Jinyu Huang}
\author [1,2,*]{Tyler R. Josephson}
\affil[1]{Department of Chemical, Biochemical, and Environmental Engineering, University of Maryland, Baltimore County, Baltimore, MD 21250}
\affil[2]{Department of Computer Science and Electrical Engineering, University of Maryland, Baltimore County, Baltimore, MD 21250}
\affil[*]{Corresponding author. Email: tjo@umbc.edu}
\date{}
\maketitle

\begin{abstract}
Chemical reactors are dynamic systems that can be described by systems of ordinary differential equations (ODEs). Reactor safety, regulatory compliance, and economics depend on whether certain states are reachable by the reactor, and are generally assessed using numerical simulation. In this work, we show how differential dynamic logic (dL), as implemented in the automated theorem prover KeYmaera X, can be used to symbolically determine reachability in isothermal chemical reactors, providing mathematical guarantees that certain conditions are satisfied (for example, that an outlet concentration never exceeds a regulatory threshold). First, we apply dL to systems
whose dynamics can be solved in closed form, such as first-order reactions in batch reactors, proving that such reactors cannot exceed specified concentration limits. We extend this method to reaction models as
complex as Michaelis-Menten kinetics, whose dynamics require approximations or numerical solutions. In all cases, proofs are facilitated by identification of invariants; we find that conservation of mass is both a principle proved from the ODEs describing mass action kinetics as well as a useful relationship for proving other properties. Useful invariants for continuous stirred tank reactors (CSTRs) were not found, which limited the complexity of reaction networks that could be proved with dL. While dL provides an interesting symbolic logic approach for reachability in chemical reactions, the bounds we obtained are quite broad relative to those typically achieved via numerical reachability analyses.
\end {abstract}
\section{Introduction}
Chemical reactors are the heart of a chemical plant in which low-value feed inputs are converted to high-value product outputs \cite{rawlings2002chemical}. Considering safety issues in reactors is necessary and equally challenging as it is vital in reactor design and control. Accidents can lead to fires and explosions, releases of toxic gases, or less dramatic emissions of chemicals that can potentially exceed regulatory limits. At least 14\% of major accidents are caused by chemical reactors\cite{kidam2013analysis}. In bioreactors, examining the results of enzyme kinetics is critical for modulating biochemical reaction rates through catalysis. Playing an essential role in metabolism, signal transduction, and cell regulation, they have been used across many diverse growing fields, including drug development, biofuel production, and food processing. Proper control systems for controlling concentration, temperature and pressure is critical for safe, compliant, and economical operation\cite{mitra2021bioreactor}.

Material and energy balances of reactors lead to linear and nonlinear systems of ordinary differential equations (ODEs). These equations describe the species concentration, temperature, and pressure over time, but how would we prove that they never exceed a given threshold? 

One of the approaches to prove the safety of a system is that computing the reachable set of states that a system can reach in (in-)finite time from the initial states. If that reachable state does not meet the unsafe regions, the safety of a system can be guaranteed. Reachability analysis has many useful applications in safety verification and fault detection \cite{lin2008fault}, quality assurance \cite{huang2002quantitative}, state and parameter estimation \cite{jaulin2002nonlinear, kieffer2006guaranteed}, uncertainty propagation\cite{harrison1977dynamic}, controller synthesis\cite{lygeros1999controllers}, and global dynamic optimization \cite{singer2006global}. An ODE is safe if its continuous solutions from a given initial conditions always stay in safe states, and it is reachable if its solutions finally enter a goal region from its initial conditions. Many studies use numerical solutions to check the reachability of a system \cite{althoff2010computing, scott2013bounds, tulsyan2017interval, tulsyan2017interval2, tulsyan2017interval3, harwood2013bounds, tulsyan2016reachability}. Prior work using bounding methods is numerical. However, we can’t use this method all the time because all systems of ODEs cannot be integrated nor solved with Laplace transforms. In such cases, numerical integration can produce solutions for specific values of constants and initial conditions, but numerical solutions are not easily extended to other instances. For example, Michaelis-Menten kinetics includes four differential equations that have no known closed-form analytical solution. To simplify this system into a single, solvable ODE, steady-state or quasi-equilibrium approximations are often made\cite{bispo2011extending, eilertsen2022validity}. However, such approximations do not hold across all times or initial concentration ranges. The method presented in this paper differs from bounding methods, we are exploring methods that use symbolic logic to check the reachability of a system by taking reaction invariants into consideration. It can prove the properties of systems of ODEs without requiring a closed-form solution, either by integration or Laplace transforms. 
Deductive reasoning, as a proof-based verification method which their logical foundations guarantee trustworthiness of reaching results is used to verifying ODE safety and reachability properties. We use differential dynamic logic (dL), a logic for deductive verification with well-established semantics axiomatic foundations, which ensure the correctness of resulting proofs. The method is applied for several case studies, and it was shown that the KeYmaera X can successfully prove the reachability of a system. Moreover, proofs in dL  using an automated theorem prover are guaranteed to be valid. KeYmaera X is a theorem prover for differential dynamic logic (dL).\cite{platzer2012complete, platzer2008differential, platzer2007differential}. In this study, we aim to utilize the advantages of KeYmaera X to verify the behavior of isothermal batch reactors described through ODEs. 

\section{Background}
Process control enables safety and efficiency during the operation of chemical and biological processes. Typically, simulation is used to predict and verify the safety and performance of a given control system, specify model parameters, and initialize conditions in order to integrate the system of ODEs forward in time and evaluate performance. However, such tests are not guaranteed to apply beyond the conditions of the simulation.

Statistical and reinforcement learning techniques have increased automation and leveraged “big data” to improve chemical process performance\cite{hoskins1992process,hussain1999review, nian2020review}. However, in the context of industrial control, this black-box, “self-driving” model comes with safety risks. An autonomous agent motivated to learn and optimize must never explore unsafe conditions \cite{garcia2015comprehensive}. While applications of the automated reasoning aspects of artificial intelligence in the chemical process industry peaked in the 1990s with expert systems\cite{rich1987model}, formal verification is a stronger alternative for simulation of safety verification as it offers logical guarantees that statistical learning approaches do not.

Formal verification methods have been applied to control algorithms as well. For example, verified control algorithms have been designed for the European train control system \cite{platzer2008differential}, automobile adaptive cruise control \cite{loos2011adaptive}, airplane collision prevention \cite{ghorbal2014hybrid}, and force feedback from a surgical robot\cite{abate2016proceedings}. 
More recently, these methods have been extended towards reinforcement learning algorithms\cite{fulton2018safe}, ensuring that autonomous robots can safely explore and learn in a factory setting given a formal guarantee that no collision occurs \cite{huisman2021formal}. Within this space, proportional-integral-derivative (PID) control (applied to automotive cruise control) has been formalized in dL \cite{arechiga2012using}. dL has also enabled verification of thermostat temperature control\cite{liebrenz2018deductive}and tank filling overflow protection \cite{ishii2013inductive}. However, little work has explored the use of dL to verify chemical reaction systems \cite{figueiredo2017applying,bohrer2022chemical}. 
Figueiredo, Martins, and Chaves apply this to biological reaction networks\cite{figueiredo2017applying} primarly focusing on models represented as piecewise linear systems.

We started this work in Fall 2021; independently, Rose Bohrer was also exploring dL for verification of chemical reactors, which was published in a formal methods conference \cite{bohrer2022chemical}. Bohrer considered irreversible reactions under adiabatic conditions in the presence of some control, the most critical reactor scenario involving runaway reactions. However, to simplify the system to accommodate the mathematics available to the theorem prover, she linearized the energy balance using a Taylor Series, and implemented ``bang-bang'' control, which instantly shuts off and quenches the reactor (a rather atypical control scenario). Nonetheless, this seminal work illustrates the opportunities for formal methods in the chemical sciences and engineering, and we have had several fruitful conversations with Bohrer after discussing her preprint.

Our work aims to present the subject for a chemical engineering audience unfamiliar with formal methods and theorem proving. We also crafted our problems with more practical chemical engineering scenarios in mind, with our first goal to explore how this approach might scale toward more complex reactions than Bohrer considered (A $\rightarrow$ B), to include series, parallel, and reversible reactions, up to the complexity of Michaelis-Menten kinetics.


\subsection{Differential Dynamic Logic}

KeYmaera X is a theorem prover for hybrid systems. Hybrid systems are dynamic systems that include interaction between discrete and continuous dynamics \cite{platzer2018logical}.

Understanding differential equations is essential to reason about hybrid systems. In fact, verifying properties of differential equations is an important component of verifying hybrid systems\cite{platzer2012structure}

KeYmaera X has a small trusted kernel of just 1700 lines of code, which includes a list of sound differential dynamic logic (dL) for verifying the properties of hybrid systems \cite{felty2015automated}. It provides techniques as tactics to check for sound reasoning, which is useful for locating bugs within a system. As mentioned earlier, It has been applied for systems such as safety-oriented autonomous vehicle control\cite{abhishek2020formal}, aircraft collision avoidance\cite{jeannin2017formally}, the European train control system\cite{platzer2009european}, and surgical robots\cite{fulton2015keymaera}. Formulas of differential dynamic logic (dL) are described in Table~\ref {Table 1} .

\begin{table}[ht]

\centering
\begin{tabular}{@{}ll@{}}
\toprule
\textbf{Formula of dL } & \textbf{Meaning}           \\ \midrule
$e=\tilde {e}$             & $e$ is equal to $\tilde {e}$   \\
$\neg P$       & true when P is not true \\
$P \land  Q$    & true when both P and $Q$ are true \\
$P \lor  Q$       & true when either P or $Q$ is true  \\
$P \leftrightarrow Q$  & true when both sides are true or false        \\
$P \to Q$                & true when P is true or $Q$ is false            \\
$[\alpha]P$  & P is true for all solutions to $\alpha$        \\
$\langle \alpha \rangle  P$                & P is true for at least one solution to  $\alpha$          \\ \bottomrule
\end{tabular}
\caption{Formulas of dL and meaning \cite{platzer2018logical}}
\label{Table 1}
\end{table}

\section{Chemical Reactions}
Consider a first-order irreversible reaction ($A\to B$). At each temperature, the rate of product $B$ is dependent on the concentration of reactant $A$ . The evolution of the concentration of B depends on the concentration of A. If the concentration of A is $C_{A}=0$, then the concentration of B does not change at all. If $C_{A}>0$, then $C_{B}$ always increases. 
The concentration of reactant $A$ is also subject to change all throughout the process $\frac{dC_{A}}{dt}=-kC_{A}$, and its change is dependent on its concentration. The concentration of reactant $A$ changes over time, which, in turn, change the concentration of product B. Overall, any change in either concentration can be followed by these two differential equations. $\frac{dC_{A}}{dt}=-kC_{A}$, $\frac{dC_{B}}{dt}=kC_{A}$.
Consider $C_{A0}$ and $C_{B0}$ as initial concentration of A and B, respectively. Solving the system of differential equations, given the following concentration profiles $C_{A} = C_{A0} exp(-kt)$ and $C_{B}=C_{B0}+C_{A0}(1-exp(-kt))$. 
The solution is much more complex for these differential equations, while a relatively simple differential equation can easily illustrate the behavior of complex processes \cite{platzer2018logical}.

To demonstrate dL addressing a problem that conventional techniques cannot solve, we first introduce basic examples for batch reactors, extending this method to Michaelis-Menten kinetics.

In 1913, Leonor Michaelis and Maud Leonora Menten published a paper\cite{michaelis1913kinetik}, in which they proposed a mathematical model of the enzyme reaction. It was represented as:

$E+S \underset{k_{r}}{\stackrel{k_{f}}{\rightleftharpoons}} ES{\stackrel{k_{cat}}{\to}} E+P $. \\where $E, S, ES, P, k_{f}, k_{r}, $ and $ k_{cat} $ are an enzyme, substrate, enzyme-substrate complex, product, forward rate constant, reverse rate constant, and catalytic rate constant, respectively. Additionally, E is a regenerated original enzyme, and P is a released product.
   
   $\frac{dP}{dt} =  k_{cat}C_{ES} $ 

   $\frac{dES}{dt} = k_{f}C_{E}C_{S} - k_{r} C_{ES} - k_{cat}C_{ES} $

   $\frac{dS}{dt} = -k_{f}C_{E}C_{S} + k_{r} C_{ES} $ 
   
   $\frac{dE}{dt} =  -\frac{dES}{dt} = -k_{f}C_{E}C_{S} + k_{r} C_{ES} +k_{cat}C_{ES}$ 
   
 This system of four differential equations has no known closed-form analytical solution. These equations are solved numerically, and the integrated form of these equations is shown in Fig. \ref {fig:Michaelis-Menten Batch}. This figure shows changes in concentration of enzyme E, substrate S, enzyme-substrate complex ES and product P over time. It can be ascertained from this diagram that the concentration of complex ES is less than the sum of the initial concentrations of enzyme E and complex ES ($ ES\leq E_{0}+ES_{0} $).

\section{Results} 
We introduce differential dynamic logic (dL) in reactor engineering using default proof procedure and tactic-based method. We unpack the logic behind KeYmaera X through a short, interpretable proof of concentration limits in a batch reactor with more detailed first-order kinetics. We finally demonstrate proof of concentration limits in a batch reactor with Michaelis-Menten kinetics, a system that cannot be integrated or solved via Laplace transform without invoking approximations.
We used KeYmaera X version 5.0.2, Python 3.9.12, Matplotlib version 3.1.3, and Scipy version 1.7.3.

\subsection{Isothermal Batch Reactors}

\subsubsection{First-Order Irreversible Reactions}
\label{First-order-irreversible-reactions-batch}

\begin{figure}%
    \centering
    \subfloat[\centering ]{{\includegraphics[width=5.5cm]{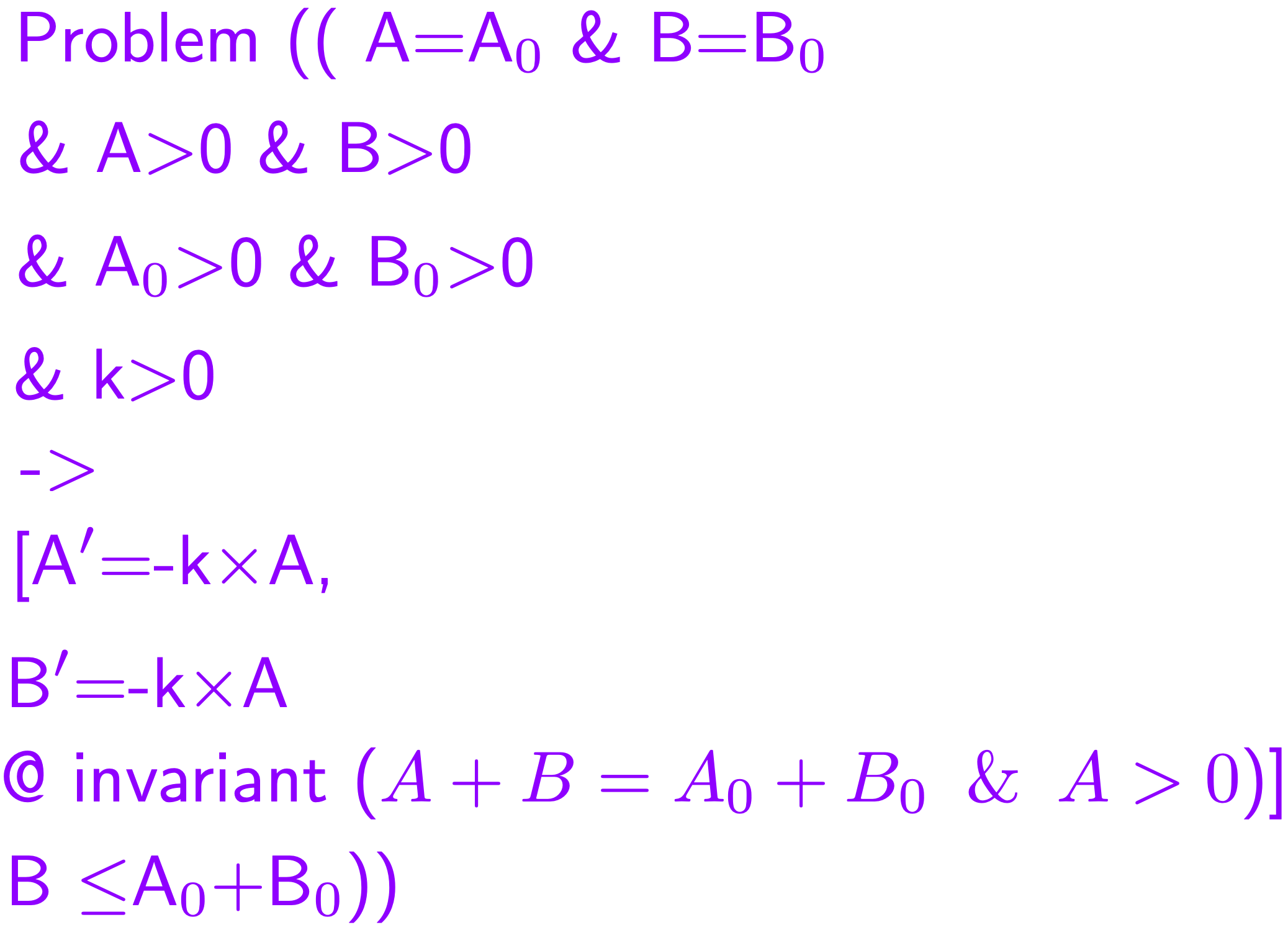} }}%
    \qquad
      \subfloat[\centering ]{{\includegraphics[width=8 cm]{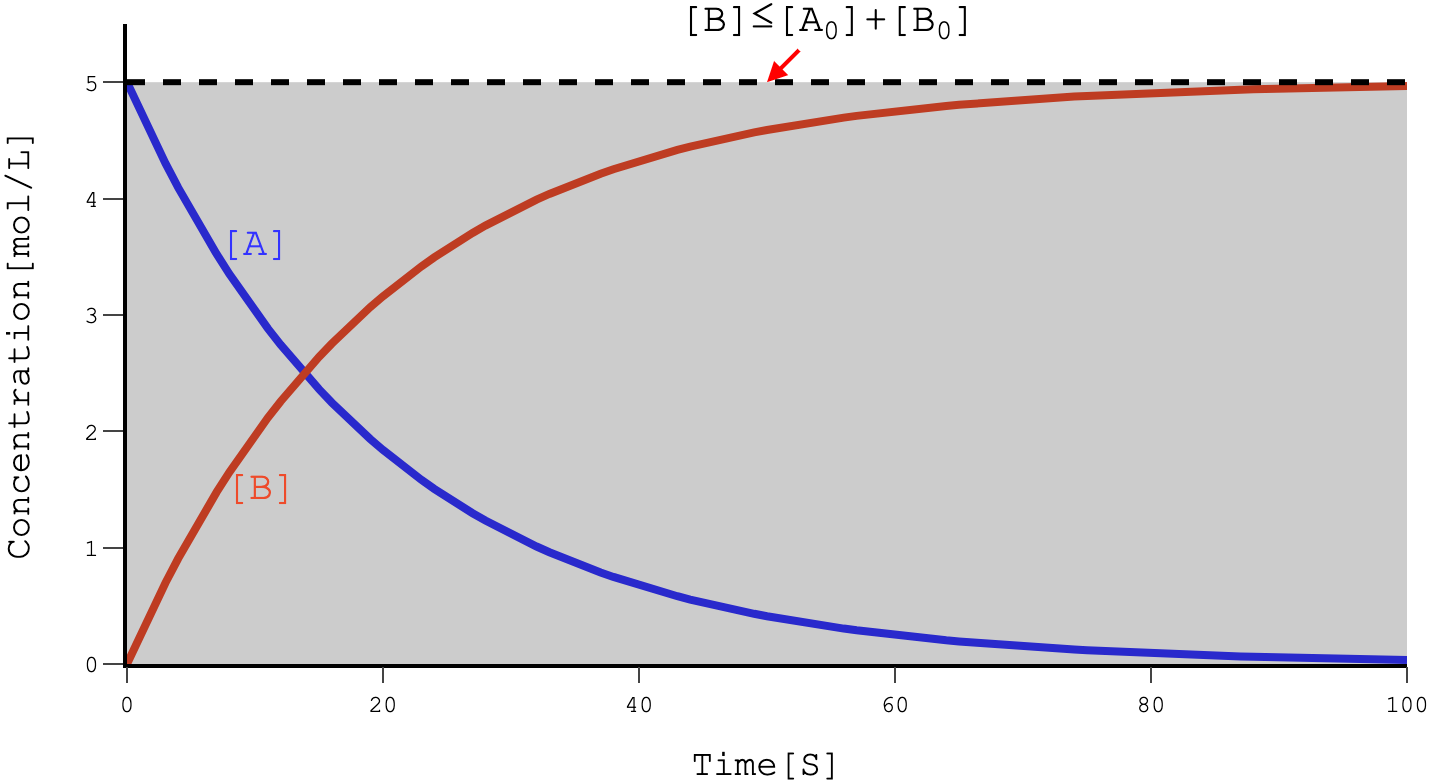} }}%
    \caption{a) The structure of proof for first-order irreversible reaction for batch reactor in KeYmaera X. b) The change in concentration of A and B with times in first-order irreversible reaction ($A{\stackrel{k_{}}{\to}} B$) for k=0.05, [A$_0$= 5\enspace $mol.L^{-1}$ and [B$_0$]= 0\enspace $mol.L^{-1}$. $[B] \leq [A_0]+[B_0]$ represents that the concentration of B at all times is less than the total initial concentration of A and B.}
    \label{fig:First-order-irreversible-KeY-integration}%
\end{figure}

We first start with first-order, irreversible reactions in batch reactors $A \overset{k} \rightarrow B$ in which we have the following: \\
For all $t>0$ where $A_{0}$ and $B_{0}$ are the initial concentrations of $A$ and $B$ respectively. Concentration of $B$ all times is less than the total initial concentration of $A$ and $B$. Fig.~\ref {fig:First-order-irreversible-KeY-integration}(b). describes the behaviour of such first order irreversible reaction in batch batch reactor. Expressed in terms of Differential Dynamic Logic (dL), our proof target can be written as:

$[\{A'=-kA, B'=kA\}]B \le A_0 + B_0$

in which $A', B' $ stands for change in concentration of A and B respectively, and $k$ is the rate constant for the reaction.  Fig.~\ref {fig:First-order-irreversible-KeY-integration}(a) shows how we can define this problem in KeYmaera X. Unfortunately, KeYmaera X takes an intractably long time to prove the above inequality without assistance. Instead, we can suggest a sub-problem for KeYmaera X to prove and then leverage it in its original proof. This is done through "Differential Cuts". The ``differential cut" rule allows you to introduce a new fact or auxiliary invariant about the continuous evolution of a system. This new fact is considered to be valid throughout the entire process of the system's evolution. The introduced fact needs independent validation, then it can be used to help simplify the proof of the main property one is interested in.

In words, we seek to introduce a restriction on the evolution domain of our system of differential equations to make it easier  in proving our original target $[B]\le [A_0]+[B_0]$. If we can also prove that the restriction necessarily results from the evolution of the system, we have proven $[B]\le [A_0]+[B_0]$ for the original (unrestricted) system. 
So, we use the differential cut rule (dC) to introduce mole balances, $[A] +[B] = [A_0] + [B_0]$ , assisting the main proof. This auxiliary ... must be prove first then we can use it to prove the main proof. To prove  $[A] +[B] = [A_0] + [B_0]$, differential cut rule is used. This rule allows you to confirm the constant truth of a specific condition throughout the evolution of a system. Since in this system this invariant as mole balence $[A] +[B] = [A_0] + [B_0]$ is true all through the system. $[A] +[B] = [A_0] + [B_0]$ is trivially true at the beginning of the system's evolution since $[A]=[A_0]$ and $[B]=[B_0]$ at $t=0$. To confirm it is true all though the system differential invariant rule is used. 
According to differential invariant rule to confirm that $A + B = A_0 + B_0$ stays constant throughout a system's evolution the derivative of $(A + B = A_0 + B_0)$ must remain true. Since $[A_0]$ and $[B_0]$ are both constants, $[A'] + [B'] = 0)$. Moreover, here we need to substitute $ [A']= -k[A]$ and $ [B']= k[B]$, in this regard Dassignb is used. Dassignb tactic relates to assigns the value of a particular expression to a variable. So, one can use ``Dassignb'' to substitute $[A']$ and $[B']$ values into $[A']+[B']=0$ which leads to $-k[A]+k[B]=0$, which indicates that the rate of change for $[A]$ and $[B]$ are the same over time, and it remains verified at all times. It is clear that $[A]+[B]= [A_0]+[B_0]$ implies $[B] \le [A_0] + [B_0] $ if $[A]>0$, so we aim to demonstrate that $[A]>0$ for all times $t>0$. This can be done with another Differential Cut, one that uses $[A]>0$ as our restriction in differential cuts. Although it may be immediately obvious that $[A]>0$ for all points in $[A']=-k[A]$ for some initial positive value of $[A]$, KeYmaera X cannot solve the differential equation directly because it involves a natural exponent in its solution. Instead, we must work around solving the equation by making use of a "Differential Ghost". Differential ghost is used to help in your analysis of the system. This auxiliary variable doesn't change the evolution domain, but it can help to derive additional insights which can in turn simplify the process of proving the property. To prove $A > 0$, we introduce a new variable $y$ into our system of equations and seek to prove a new post condition $Ay^2=1$. If this new post condition implies our original one, then we have proved that the original was trivially true in the original system. After the application, we are once again faced with two statements that need to be proved:

\begin{figure}%
    \centering
    \subfloat[\centering ]{{\includegraphics[width=16.5cm]{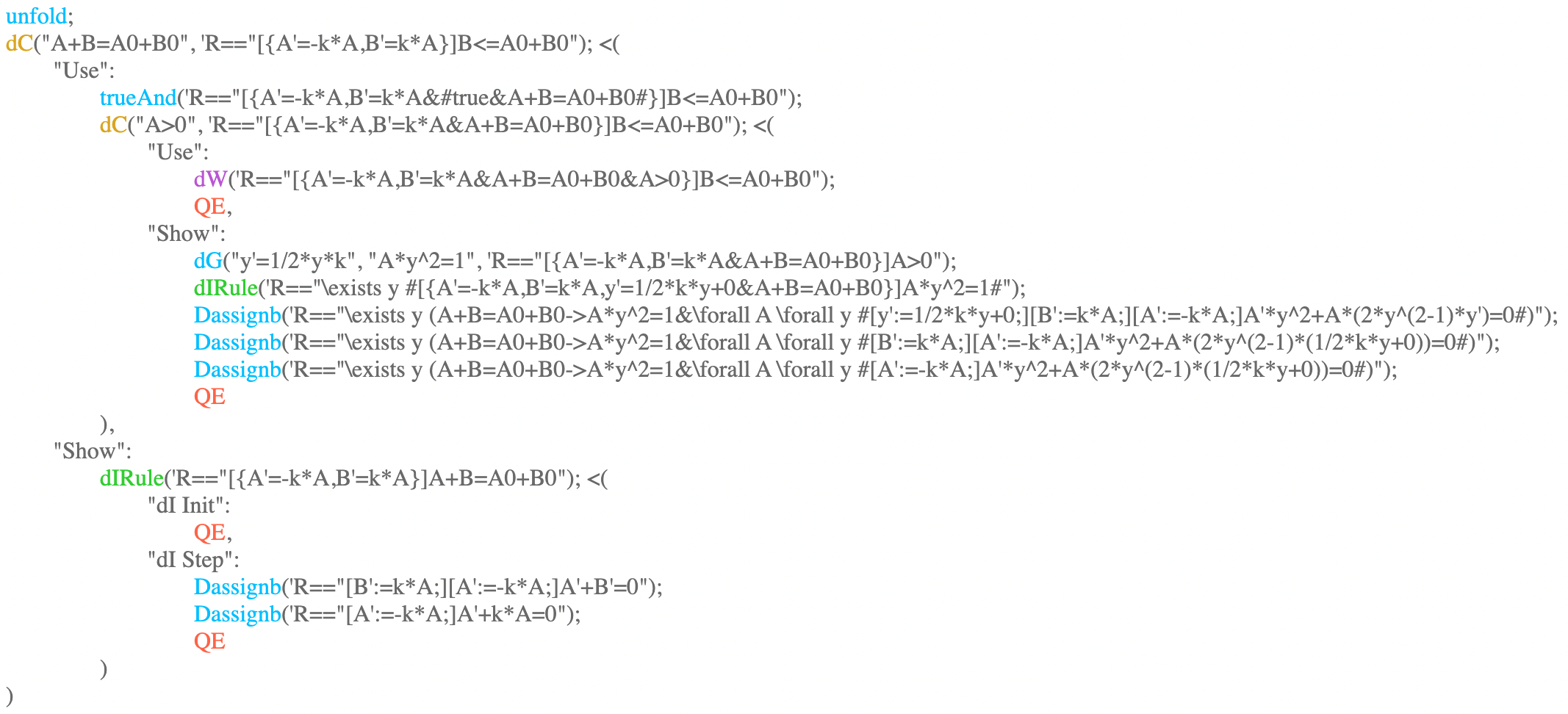} }}%
    \qquad
      \subfloat[\centering ]{{\includegraphics[width=11 cm]{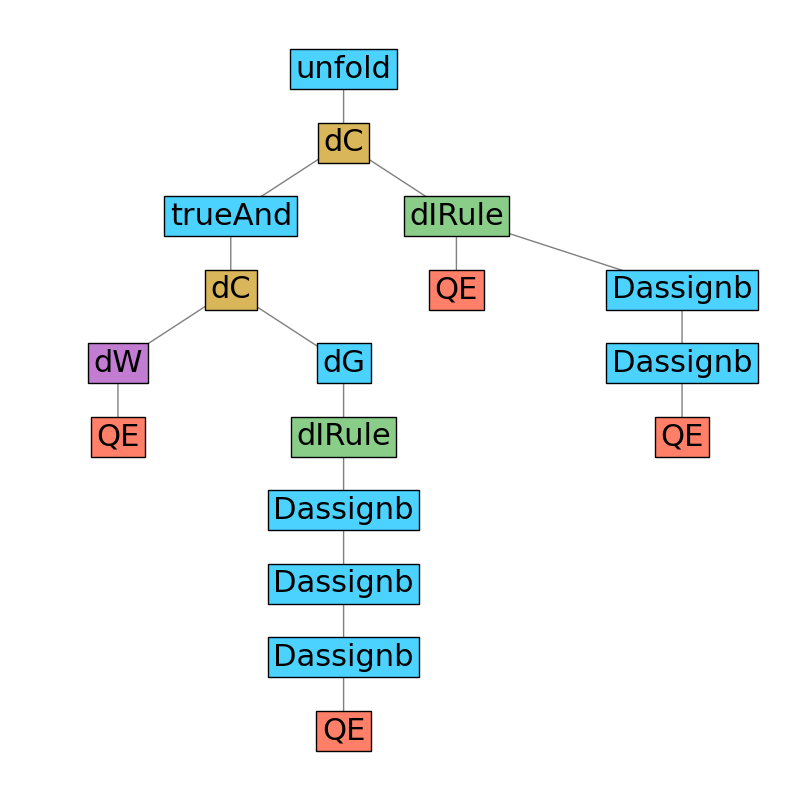} }}%
    \caption{a) full proof of first-order irreversible reaction in batch reactor. b) Visualization of this proof as a tree}
    \label{fig:1st-order-batch-Full-Chart-proof-firt-order-batch-reactor}%
\end{figure}

$\exists y [\{A'=-kA,B'=kA,y'=\frac{1}{2}yk \; \& \; A+B=A_0+B_0\}]Ay^2=1$

and\\
$Ay^2=1\vdash A>0$

The second statement is trivially true since $y^2$ must be positive (and therefore $A=\frac{1}{y^2}$ must also be positive). The first statement can be proved using a Differential Invariant, and our proof is complete. Fig.~\ref {fig:1st-order-batch-Full-Chart-proof-firt-order-batch-reactor}(a). describes a complete proof of this reaction, and Fig.~\ref {fig:1st-order-batch-Full-Chart-proof-firt-order-batch-reactor}(b) illustrates the summary of the whole structure of proof for the first-order irreversible reaction. 

\subsubsection{Parallel and Series Reactions}
\begin{figure}%
    \centering
    \subfloat[\centering ]{{\includegraphics[width=5.5cm]{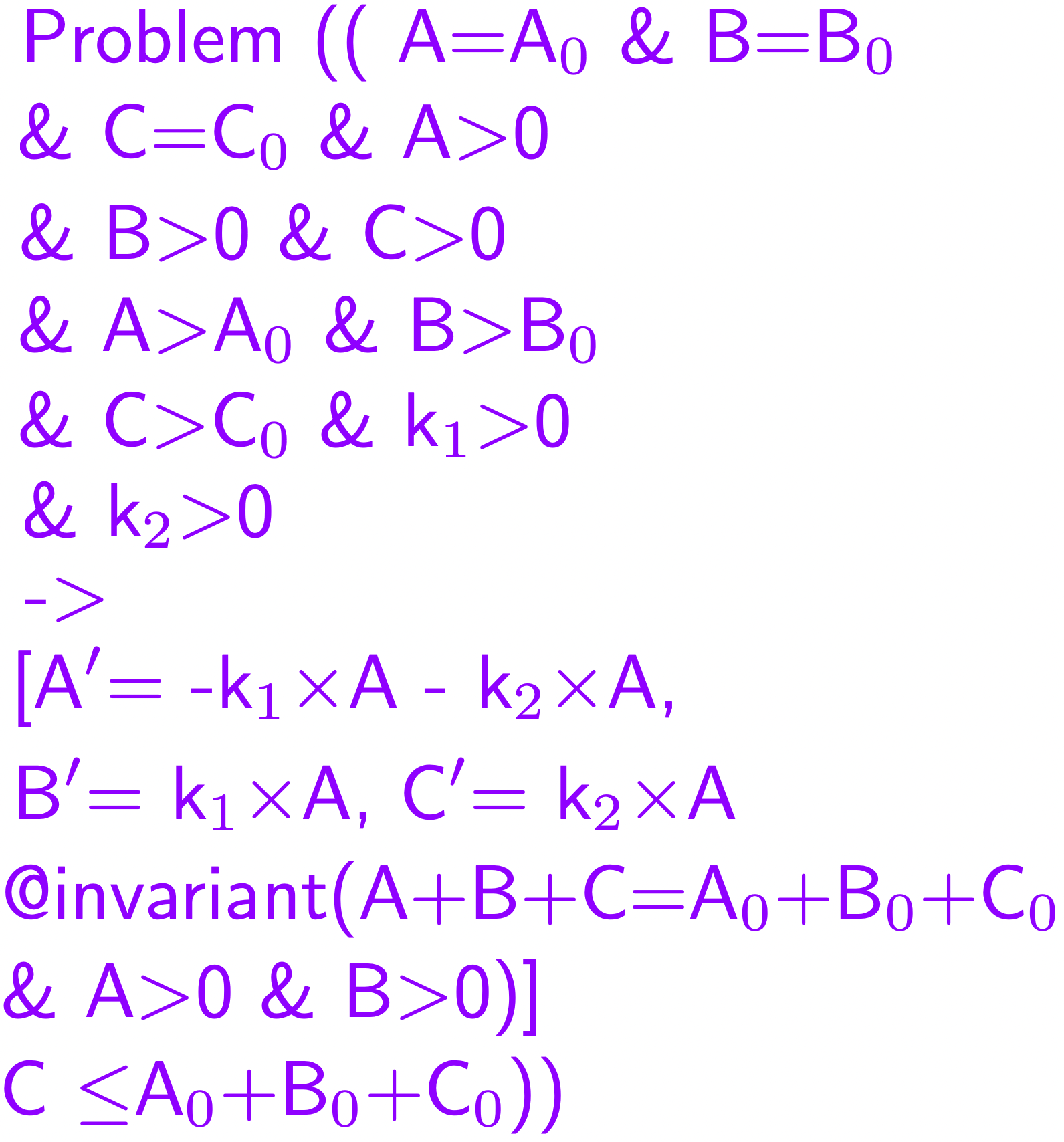} }}%
    \qquad
      \subfloat[\centering ]{{\includegraphics[width=9 cm]{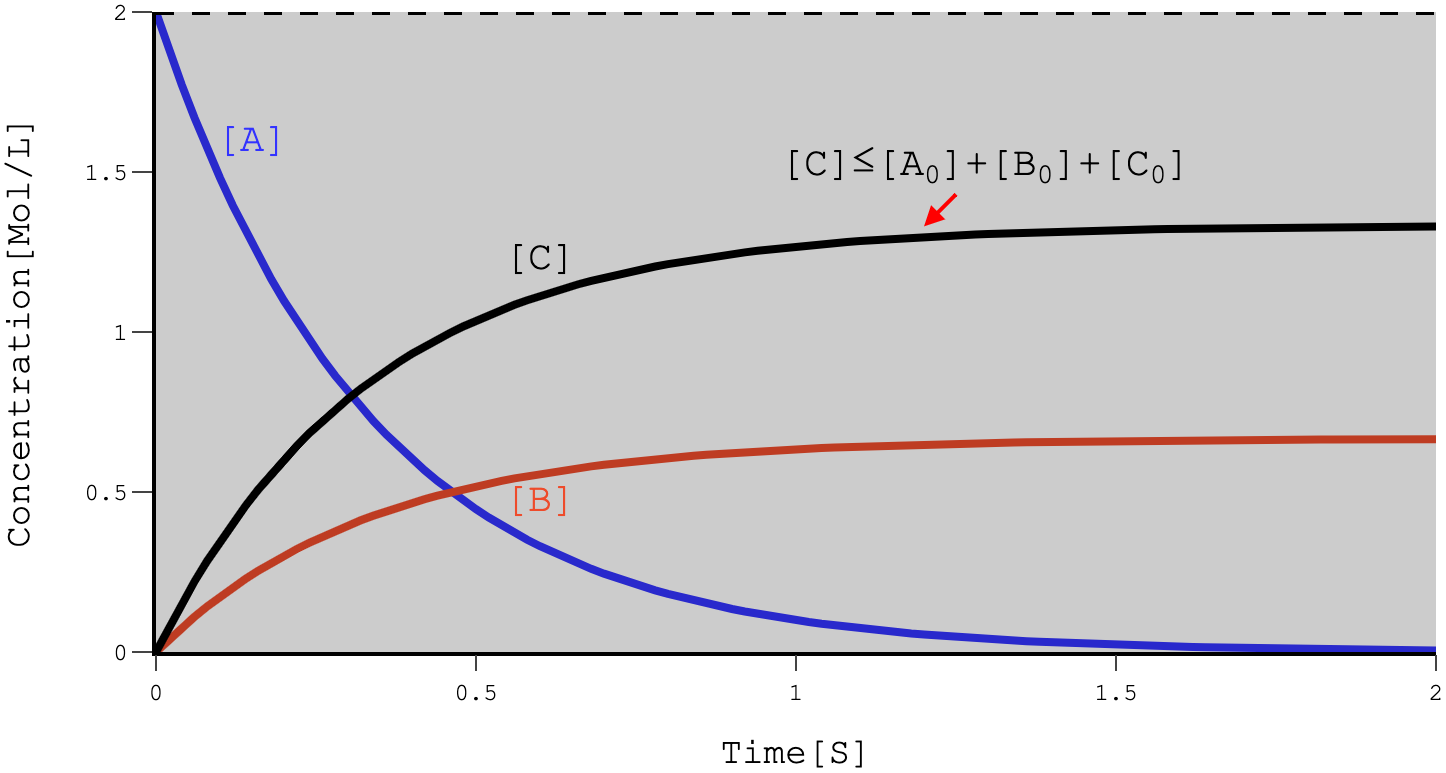} }}%
    \caption{a) KeYmaera X input file for prooving two parallel reactions in batch reactor. b) Numerical solution illustrating the changes in concentration of components in parallel reactions $A\stackrel{k_{1}}{\to}B,A\stackrel{k_{2}}{\to}C$in batch reactor for k$_1$=1, k$_2$=2, [A$_0$]=2\enspace $mol.L^{-1}$, [B$_0$]=0 \enspace $mol.L^{-1}$, and  [C$_0$]=0 \enspace  $mol.L^{-1}$. $ [C] \leq [A_0]+[B_0]+[C_0]$ shows that the concentration of C is always less than the total initial concentration of A, B, and C. Detailed proof and proof graph are in the SI}
    \label{fig:two-parallel-reaction-KeY-Integral}
\end{figure}

\begin{figure}%
    \centering
    \subfloat[\centering ]{{\includegraphics[width=5.5cm]{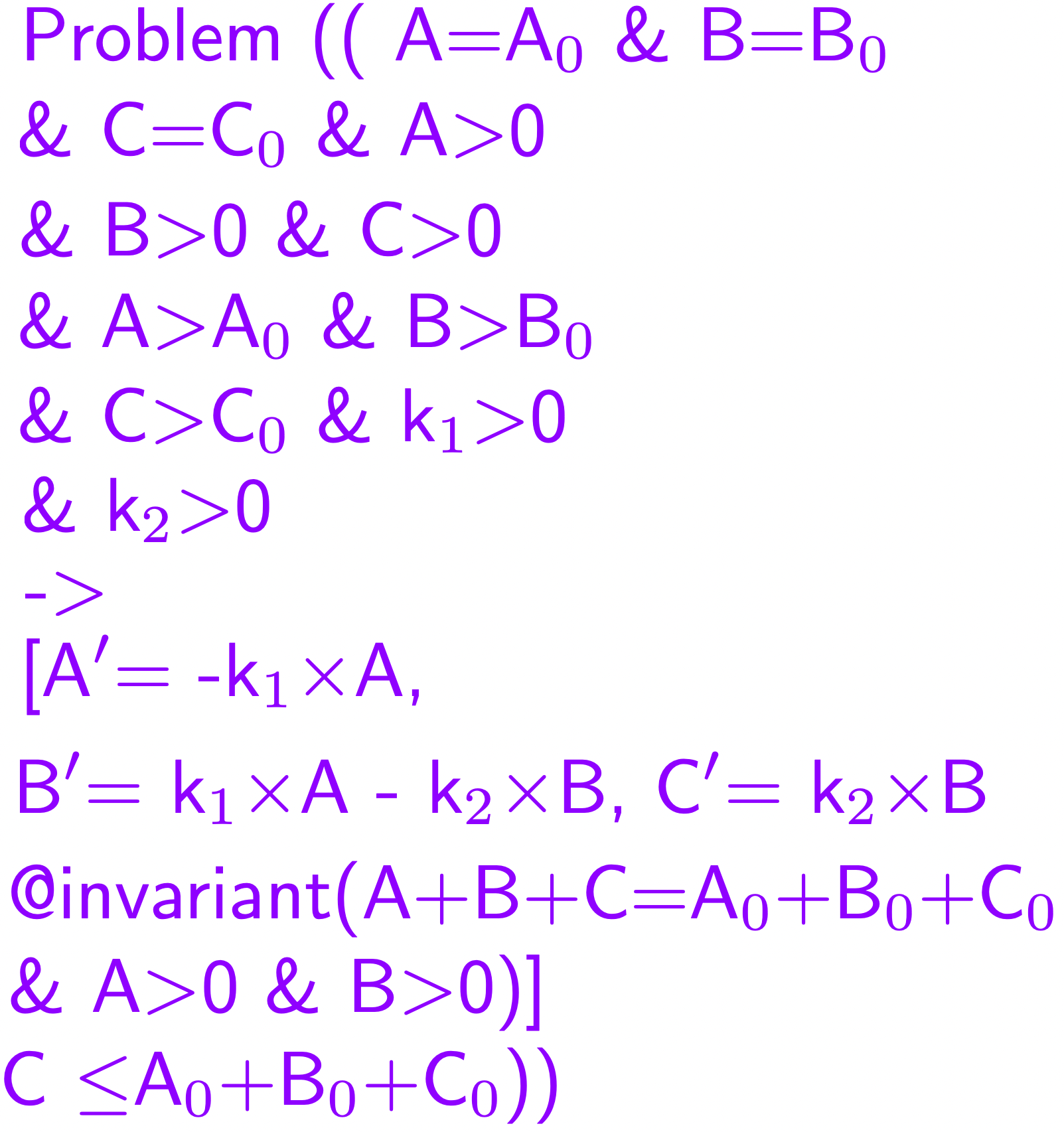} }}%
    \qquad
      \subfloat[\centering ]{{\includegraphics[width=9 cm]{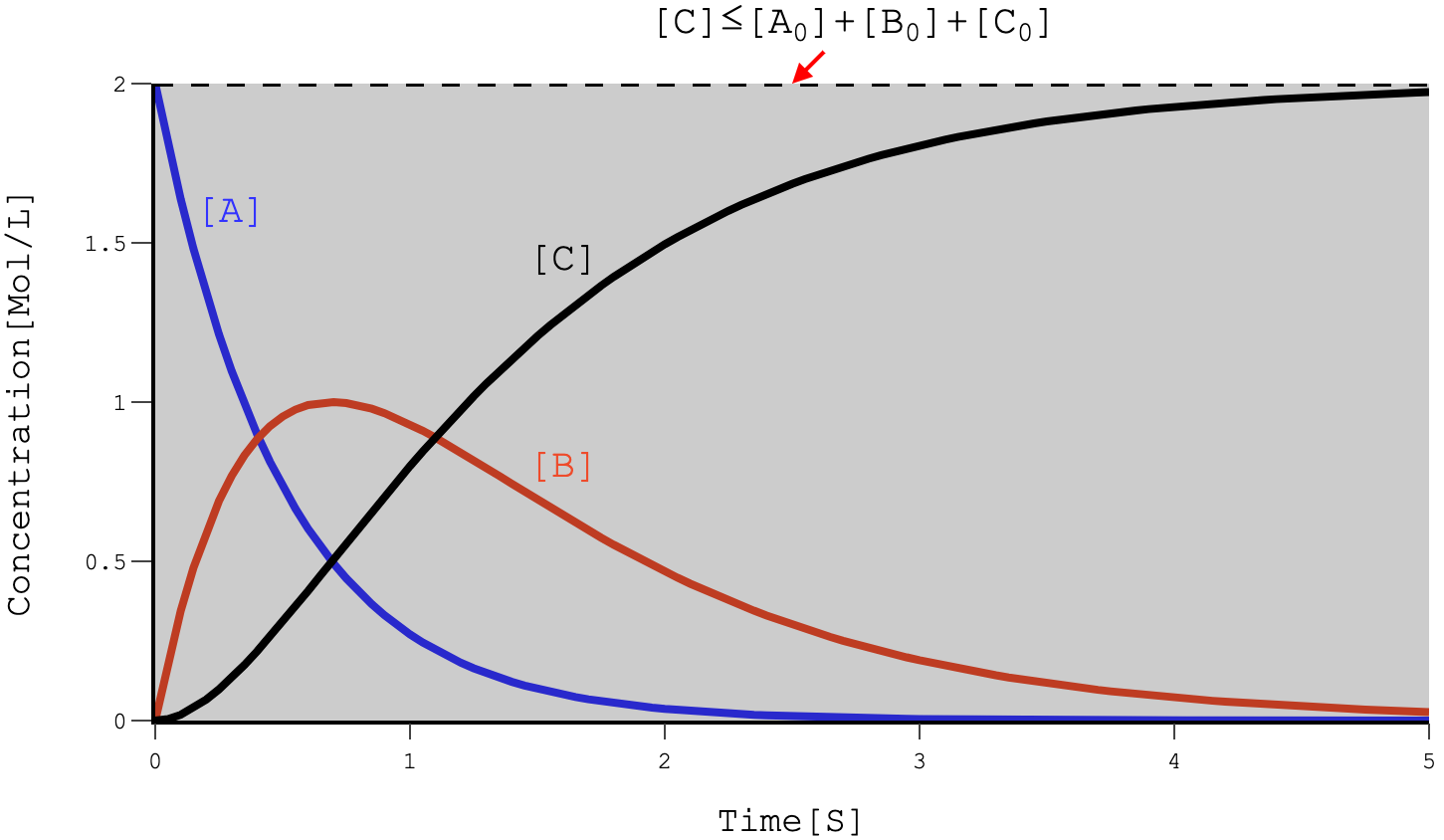} }}%
    \caption{a) KeYmaera X input file for proving two series reactions in batch reactor in KeYmaera X. b) Numerical solution illustrating the changes in concentration of components in the reaction $A\stackrel{k_{1}}{\to}B\stackrel{k_{2}}{\to}C$ with $k_{1}=2 , k_{2}=1,  [A_{0}]=2 \enspace mol.L^{-1}$, $ [B_{0}]=0 \enspace mol.L^{-1} , [C_{0}]=0 \enspace mol.L^{-1}. [C] \leq [A_0]+[B_0]+[C_0]$. Detailed proof and proof graph are in the SI. }
    \label{fig:two-series-reaction-KeY-Integral}%
\end{figure}

Moreover, we apply the same procedure for two parallel reactions  $A{\stackrel{k_{1}}{\to}B,B{\stackrel{k_{2}}{\to}C }}$ and two reactions in series  $A{\stackrel{k_{1}}{\to}B{\stackrel{k_{2}}{\to}C }}$ . (Fig.~\ref{fig:two-parallel-reaction-KeY-Integral})(b) and (Fig.~\ref {fig:two-series-reaction-KeY-Integral})(b) describe numerical solutions of these reactions. It is obviously that the concentration of $C$ in both reactions are less than the total initial concentration of $A$, $B$ and $C$ ($ [C] \leq [A_0]+[B_0]+[C_0]$). Fig.~\ref{fig:two-parallel-reaction-KeY-Integral}(a) and Fig.~\ref {fig:two-series-reaction-KeY-Integral}(a) show the definition of two parallel and two series reaction in KeYmaera X, respectively, with compare proofs available in the SI. These proofs are longer and require more steps, but can proceed using the same tactics as the proof for the first-order irreversible reaction.

\begin{figure}%
    \centering
    \subfloat[\centering ]{{\includegraphics[width=6.5cm]{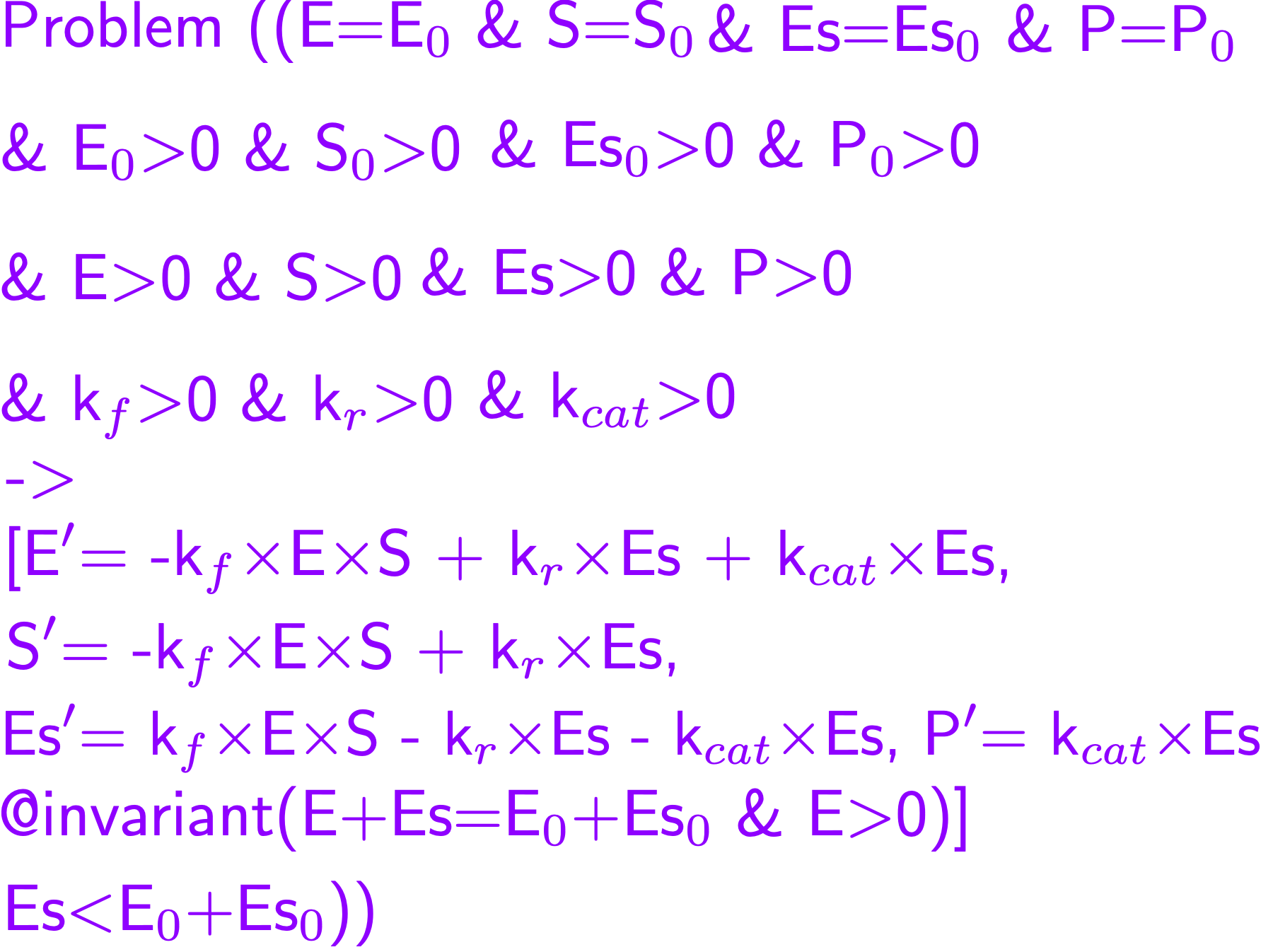} }}%
    \qquad
      \subfloat[\centering ]{{\includegraphics[width=8 cm]{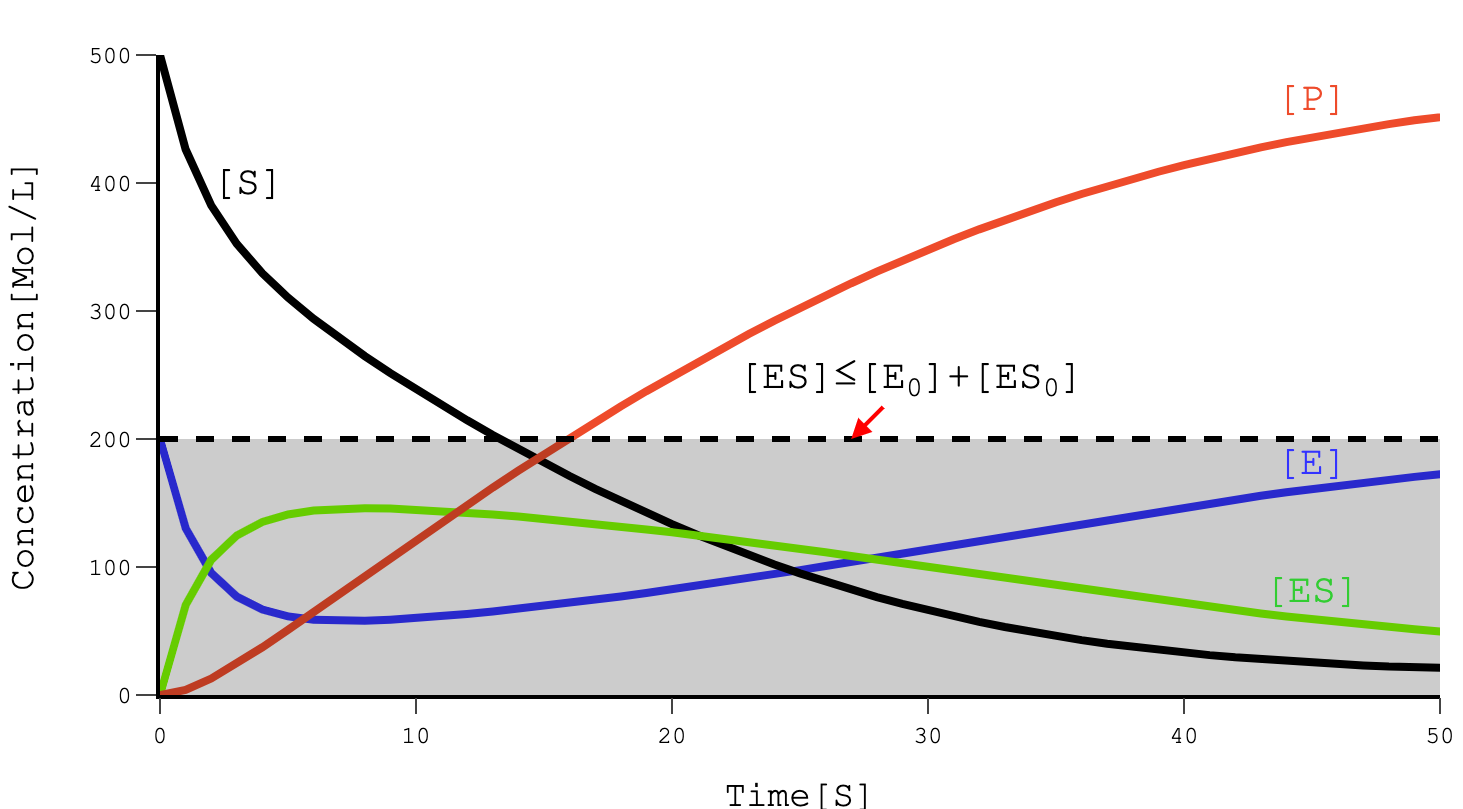} }}%
    \caption{a) KeYmaera X input file for proving [ES] < E$_0$ + ES$_0$ in Michaelis-Menten kinetics in batch reactor. b) Numerical integration for Michaelis-Menten kinetics
    $E+S \underset{k_{r}}{\stackrel{k_{f}}{\rightleftharpoons}} ES{\stackrel{k_{cat}}{\to}} E+P $ with $k_{f}= 0.001,\enspace k_{r}= 0.0001,\enspace k_{cat}=0.1,\enspace$ and initial concentration $[E_0]=200  \enspace mol.L^{-1},\enspace [S_0]=500 \enspace mol.L^{-1},\enspace [ES_0]=0 \enspace mol.L^{-1}, \enspace [P_0]=0  \enspace mol.L^{-1}$ }
    \label{fig:two-series-reaction-KeY-Integral
}%
    \label{fig:Michaelis-Menten Batch}%
\end{figure}

\subsubsection{Michaelis-Menten Reaction}
\begin{figure}%
    \centering
    \subfloat[\centering ]{{\includegraphics[width=18.5cm]{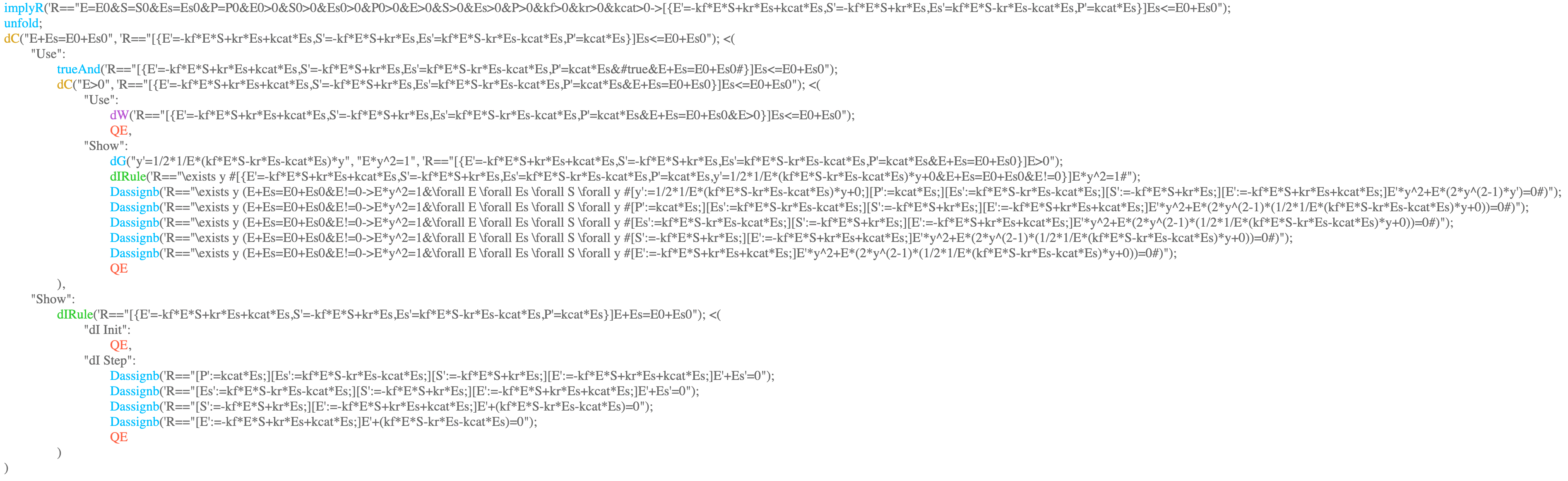} }}%
    \qquad
      \subfloat[\centering ]{{\includegraphics[width=11 cm]{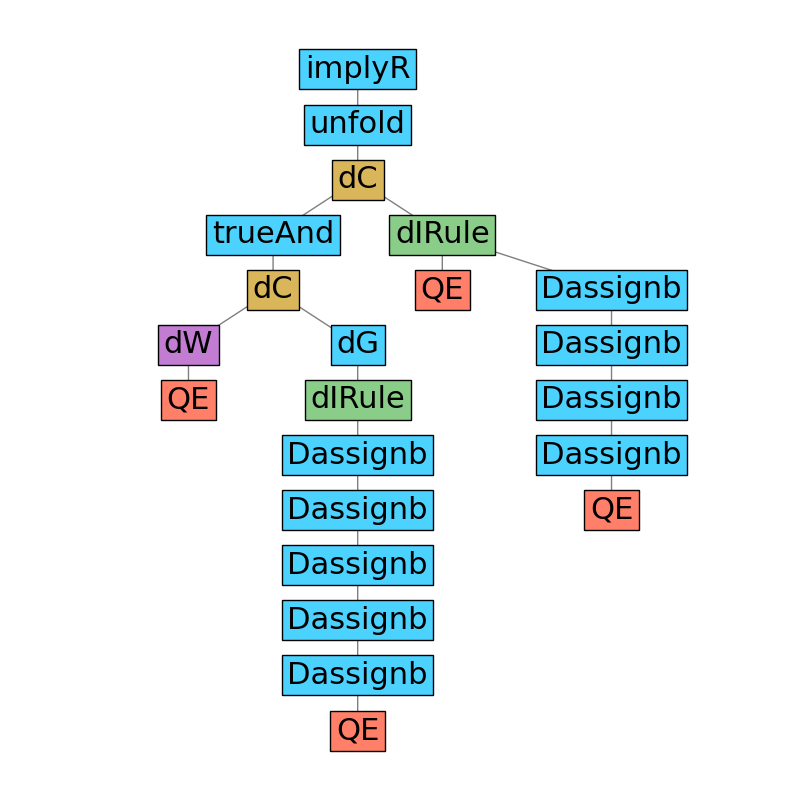} }}%
    \caption{a) full proof of Michaelis-Menten kinetics in batch reactor. b) Chart proof for Michaelis-Menten}
    \label{fig:Full-Chart-proof-Michaelis-Menten-Kinetics}
\end{figure}

Finally, to demonstrate dL addressing problems that cannot be solved by conventional techniques, we used KeYmaera X to prove the concentration properties of Michaelis-Menten kinetics \ref{fig:Full-Chart-proof-Michaelis-Menten-Kinetics}. The numerical integration solution of this reaction is illustrated in Fig.~\ref {fig:Michaelis-Menten Batch}. $ES$ is apparently less than the total initial concentration of $E$ and $ES$ ($[ES] \leq [E_0]+[ES_0]$), and this condition  holds all times. KeYmaera X can successfully verify the correctness of the condition for Michaelis-Menten kinetics.

\subsection{Isothermal CSTRs}
Next, we used KeYmaera X to prove statements about the concentration properties of first-order irreversible reactions in continuous stirred-tank reactors (CSTRs).
\subsubsection{First-Order Irreversible Reactions}
\begin{figure}%
    \centering
   {{\includegraphics[width=9cm]{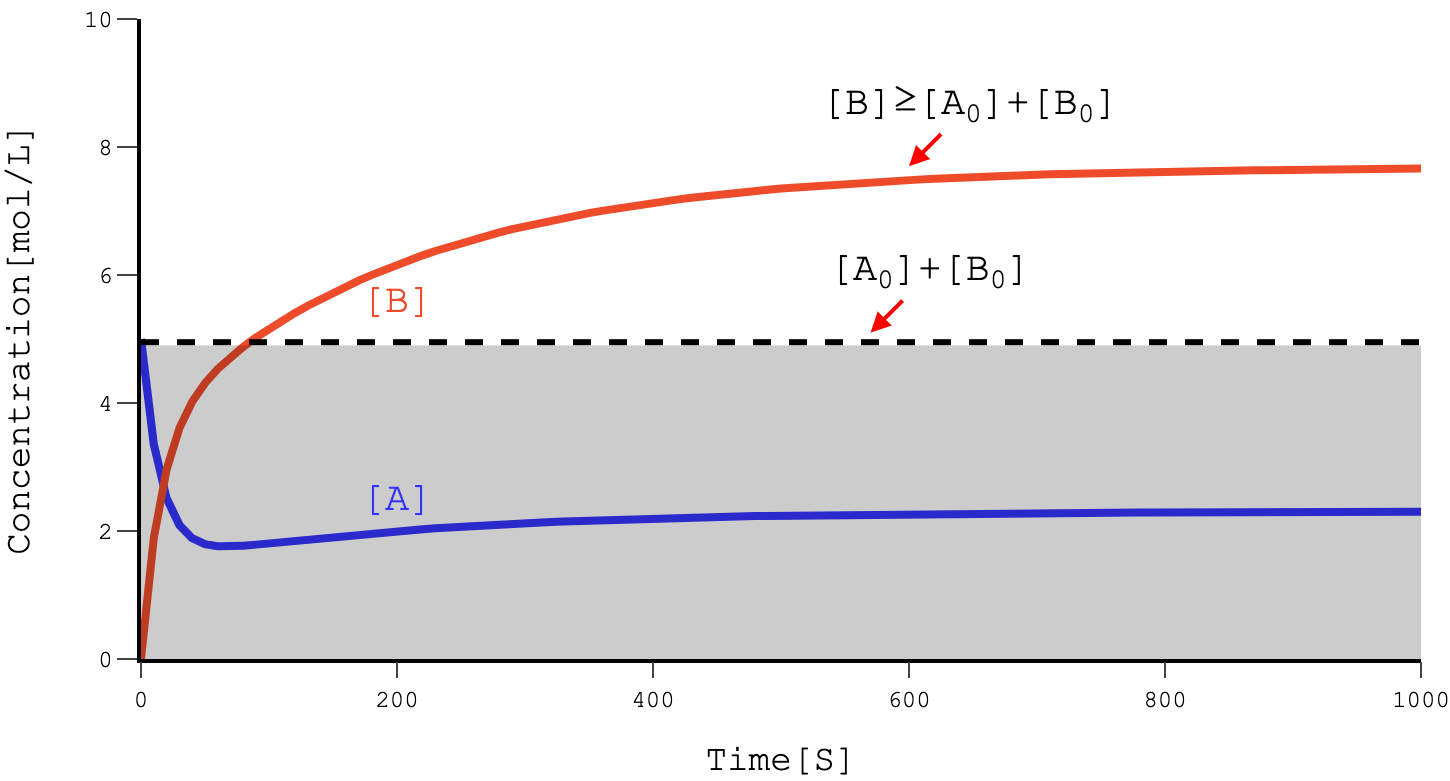} }}%
    \caption{First-order irreversible reaction$(A{\stackrel{k_{}}{\to}} B)$ in CSTR. The condition $[B] \leq [A_0]+[B_0]$ which was true for the batch reactor is not true in this case.}
    \label{fig:First-order CSTR}%
\end{figure}

While we have proven that $B \leq A_{0}+B_{0}$ in all first-order, irreversible batch reactors (Section ~\ref{First-order-irreversible-reactions-batch}), this condition is not always true in CSTRs because of the additional terms for input and output (for example,  Fig.~\ref {fig:First-order CSTR}). 
The additional terms also prevent us from identifying useful invariants, because $A + B \neq A_0 + B_0$ in a CSTR, for all values of the input flows. Consequently, we were not able to create symbolic proofs using invariants, as we did for batch reactors. This doesn't mean that no invariants exist; we consider identification of useful invariants for CSTR proofs to be a subject for future work.

\begin{figure}%
    \centering
    \subfloat[\centering ]{{\includegraphics[width=6.5cm]{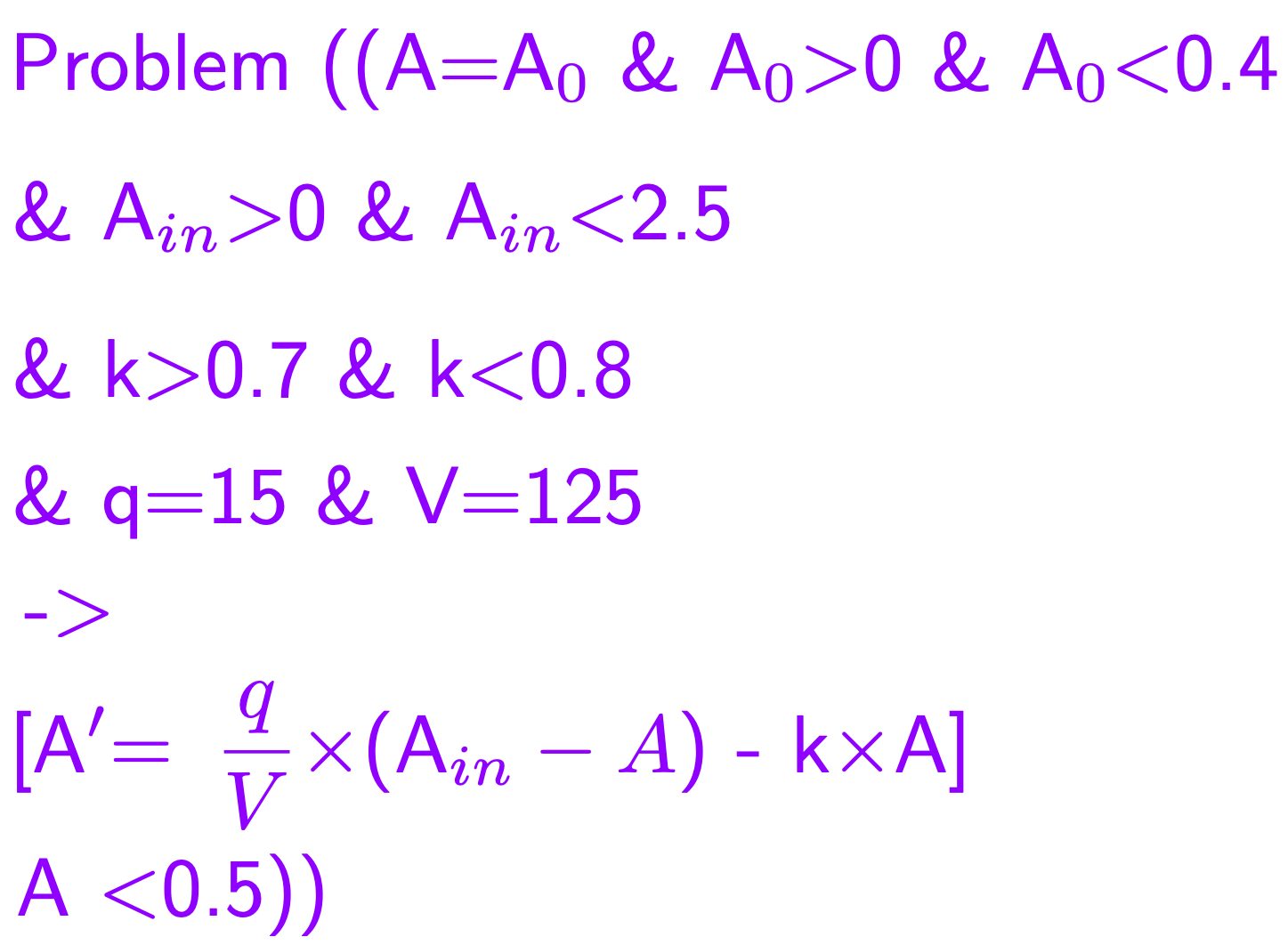} }}%
    \qquad
      \subfloat[\centering ]{{\includegraphics[width=8 cm]{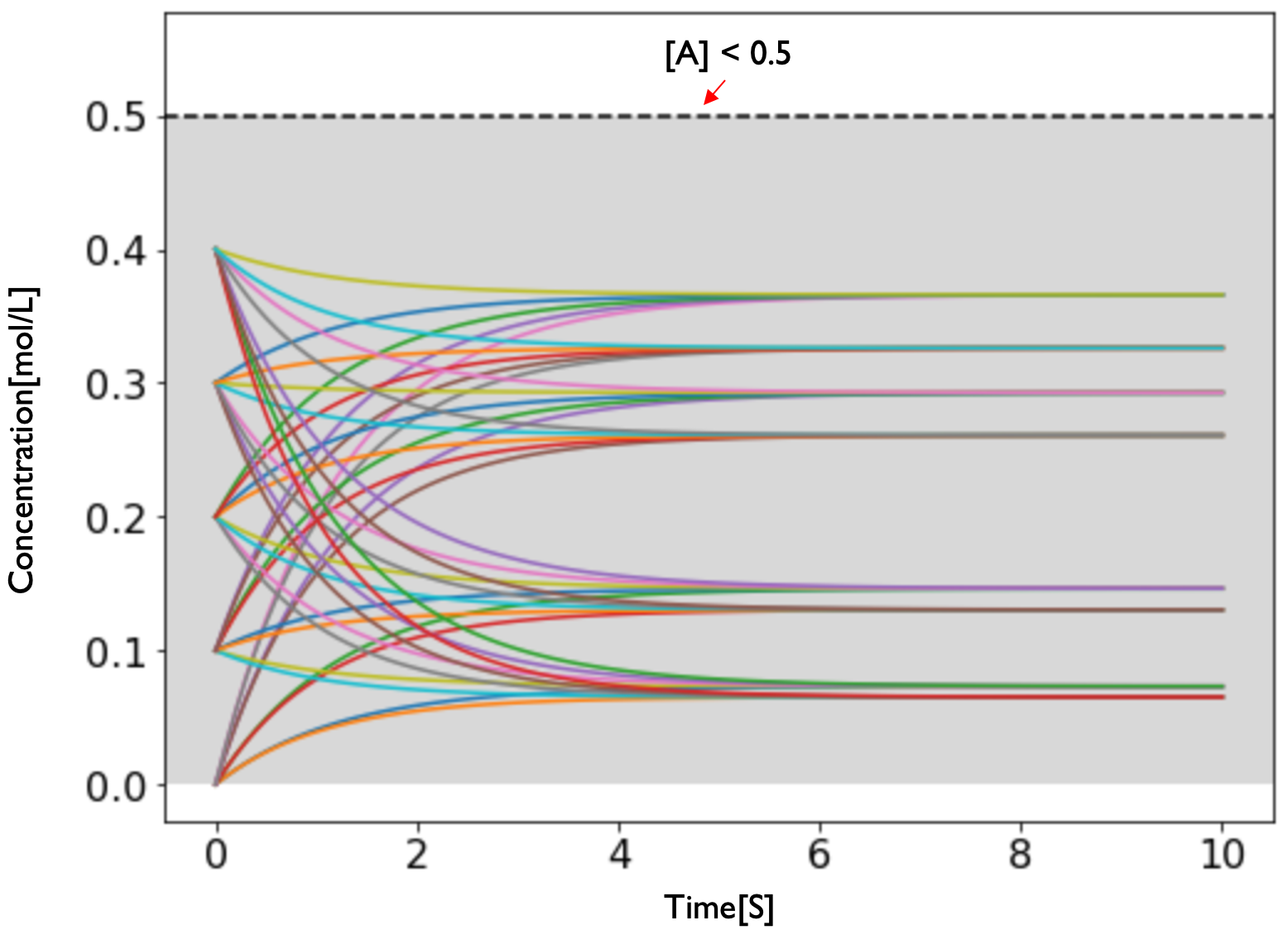} }}%
    \caption{First-order irreversible reaction $(A{\stackrel{k_{}}{\to}} B)$ in CSTR with k=0.05, \enspace A$_0$=5 $mol.L^{-1}$, \enspace B0=0 $mol.L^{-1}$,\enspace A$_{in}$=10 $mol.L^{-1}$, \enspace B$_{in}$=0 $mol.L^{-1}$, k=0.05, q=5 $L.s^{-1}$, V=1000 L.}
    \label{fig:Numerical_First_CSTR}%
\end{figure}
Nonetheless, the automated proof heuristics in KeYmaera X are capable of generating proofs for CSTRs when parameters are constrained to numerical ranges. Fig.~\ref {fig:Numerical_First_CSTR}). is an example of a first order irreversible reaction in CSTR, in which $A_{in}$ is the concentration of reactant A in the feed, k is the reaction rate constant, q is flow rate, and V is the volume of a reactor.

\section{Discussion and Outlook}

KeYmaera X is a tactic-based prover and for our simplest proofs, it could resolve them fully-automatically. Systems of ODEs with more parameters and fewer symbols were more amenable to automatic solving.

When automatic solving was not possible, what was most needed was to provide an invariant to KeYmaera X. This was the most challenging part of the proofing process. Invariants have a deep root and connection with the physics of a system.
For example, in our case study for batch reactors, we implemented an invariant to reliably prove our system and disprove others.
But the same invariant isn't applicable for CSTRs.

More complex reaction networks, where the mechanism has more than one elementary step, also led to more difficult proofs.
There has been some precedent for proving properties of large, complex systems that are comprised of smaller, modular, components, such as for distributed control of autonomous vehicles \cite{renshawDistributedTheoremProving2011}.
Reaction networks are comprised of elementary components, so scaling might be achieved if the network can be broken down and logically analyzed in parts. We found that our experience studying the series, parallel, and equilibrium reactions helped us learn about invariants that would prove useful in the Michaelis-Menten case, which includes elements of each. 
But our experiments overall indicate issues with scaling toward larger reaction networks -- we guess that to use dL to prove properties about highly interconnected networks, additional advances in proof modularity would need to be developed.

Future work could extend these systems of ODEs to include the energy balance and temperature effects, as well as address questions of stability \cite{tan2021deductive} for these systems of ODEs for chemical and biochemical system of ODEs \cite{wilhelm2009smallest}.
Bohrer previously explored adiabatic reaction conditions \cite{bohrer2022chemical}, approximating the energy balance using a Taylor Series. Future work could numerically evaluate the conditions under which this approximation is valid, as well as develop new approaches using the explicit Arrhenius dependence of reaction rate on temperature, as well as systematically considering various reactor configurations.


\section*{Conclusions}
Using formal logic, we successfully proved theorems about the reachability of some isothermal chemical reactions in batch reactors and CSTRs. Importantly, we didn't need to invoke an analytical solution to make conclusions about long-time outcomes of the chemical reactions. This allowed us to prove properties about Michaelis-Menten kinetics, without numerically integrating or adding assumptions about the reaction. That said, this approach, at the moment, doesn't appear to scale well to large reaction networks, nor are the bounds achieved particularly tight; most proofs only set broad upper bounds on concentration of products or intermediates based on total moles available.

\section*{Code availability}

KeYmaera X files for the systems presented here are available on the ATOMS Lab \href{https://github.com/ATOMSLab/VerifiedReactors}{GitHub}.


\section*{Acknowledgements}
We are grateful to Andre Platzer and Yong Kiam Tan for their guidance and useful comments in developing proofs. We are also grateful for many helpful conversations with Rose Bohrer on KeYmaera X, as well as formal verification in general. Funding was provided by UMBC startup funds. 

\bibliographystyle{unsrt}
\bibliography{references} 
\newpage
\section{Supplementary Information}

\section*{Glossary of Mathematical Terms and Symbols}\label{Glossary}

\begin{longtblr}{p{0.1\textwidth} p{0.41\textwidth} p{0.41 \textwidth}}
\hline
\textbf{Term}        & \textbf{Definition}                                                                                                                            & \textbf{Example}                                                                                                                                    \\ \hline
Theorem            & Theorem is a mathematical fact that has been proven to be true based on a set of basic truths and other already-proven theorems, using valid mathematical rules and operations.  

& Consider this example: $\forall x \in \mathbb{R}, x^2 \geq 0 $, for any real number x, when you square it, the result is either zero (in the case of x=0) or positive (for all other values of x)
\\
Proof           & A proof is a logical argument that conclusively shows the truth of a mathematical statement or theorem from accepted premises like axioms or established theorems 

& The proof of the statement $\forall x \in \mathbb{R}, x^2 \geq 0 $ uses both the Order Property and the Closure Property of real numbers to ensure that the square of any real number is non-negative.
\\
ImplyR             & ImplyR stands for implication right, a rule employed in sequent calculus to demonstrate implications. If an implication shows up on the right side of the turnstile symbol, we can utilize ImplyR to take the antecedent as a given and then work to validate the consequent (the portion following then).  

& Consider this example $\boldsymbol{\Gamma} \vdash A \to B, \boldsymbol{\Delta}$ by applying the $\to $ R rule we have  $\boldsymbol{\Gamma}, A  \enspace  \vdash B, \boldsymbol{\Delta}$ by assuming its left-hand side A in the list of all assumptions before $\vdash $ and proceeds to prove its right-hand side B  
       \\

QE & QE stands for Quantifier Elimination, which is an algorithm that reformulates expressions to remove quantifiers such as $\forall$ and $\exists$. It also reduces arithmetic expressions and inequalities with no free variables to true or false.

& The expression $\forall x \: ax+b=0$ is equivalent to $a=0 \vee b \ne 0$, and $2<3$ is equivalent to true.

\end{longtblr}

\begin{figure}%
    \centering
    \subfloat[\centering ]{{\includegraphics[width=16.5cm]{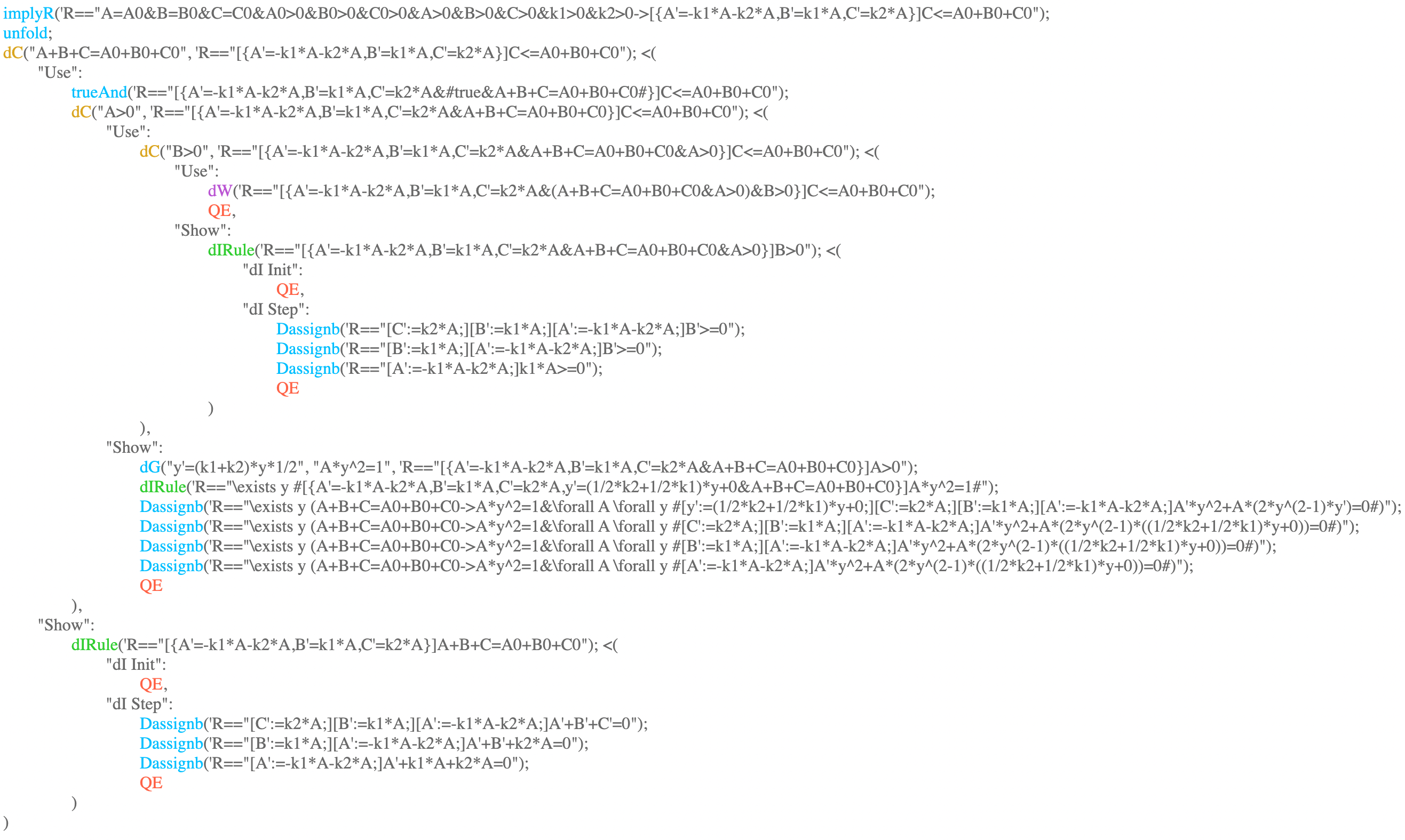} }}%
    \qquad
      \subfloat[\centering ]{{\includegraphics[width=11 cm]{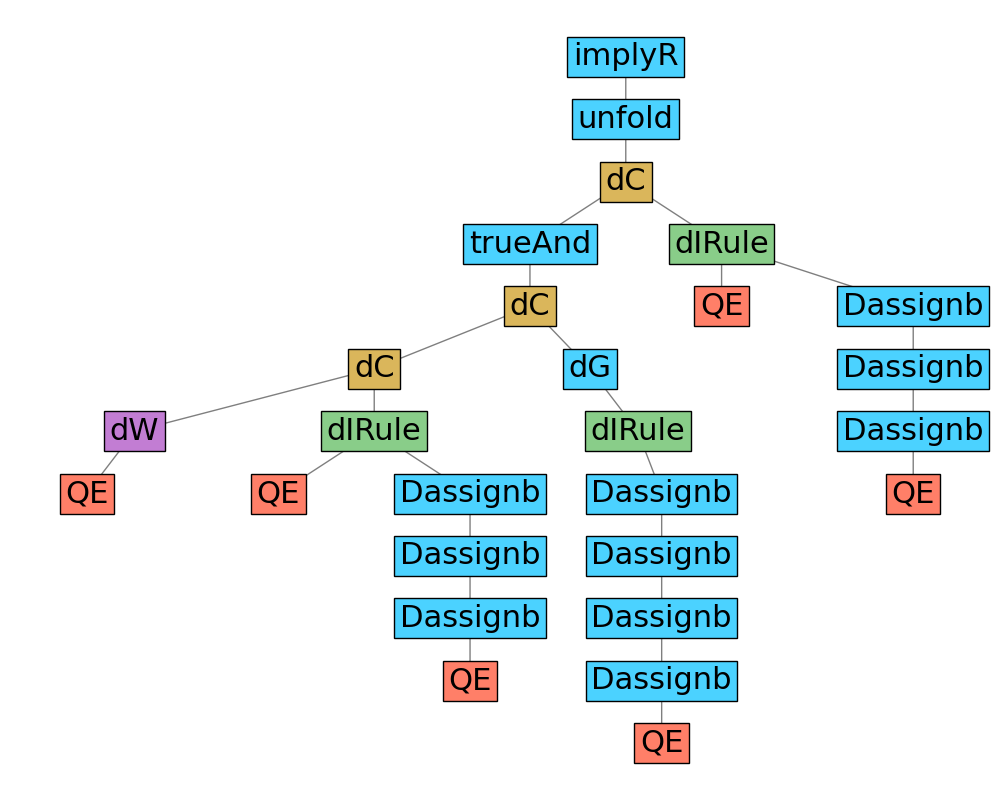} }}%
    \caption{a) full proof of two parallel reactions in batch reactor. b) Chart proof for two parallel reactions}
    \label{fig:Full-Chart-proof-parallel-reactions}%
\end{figure}

\begin{figure}%
    \centering
    \subfloat[\centering ]{{\includegraphics[width=16.5cm]{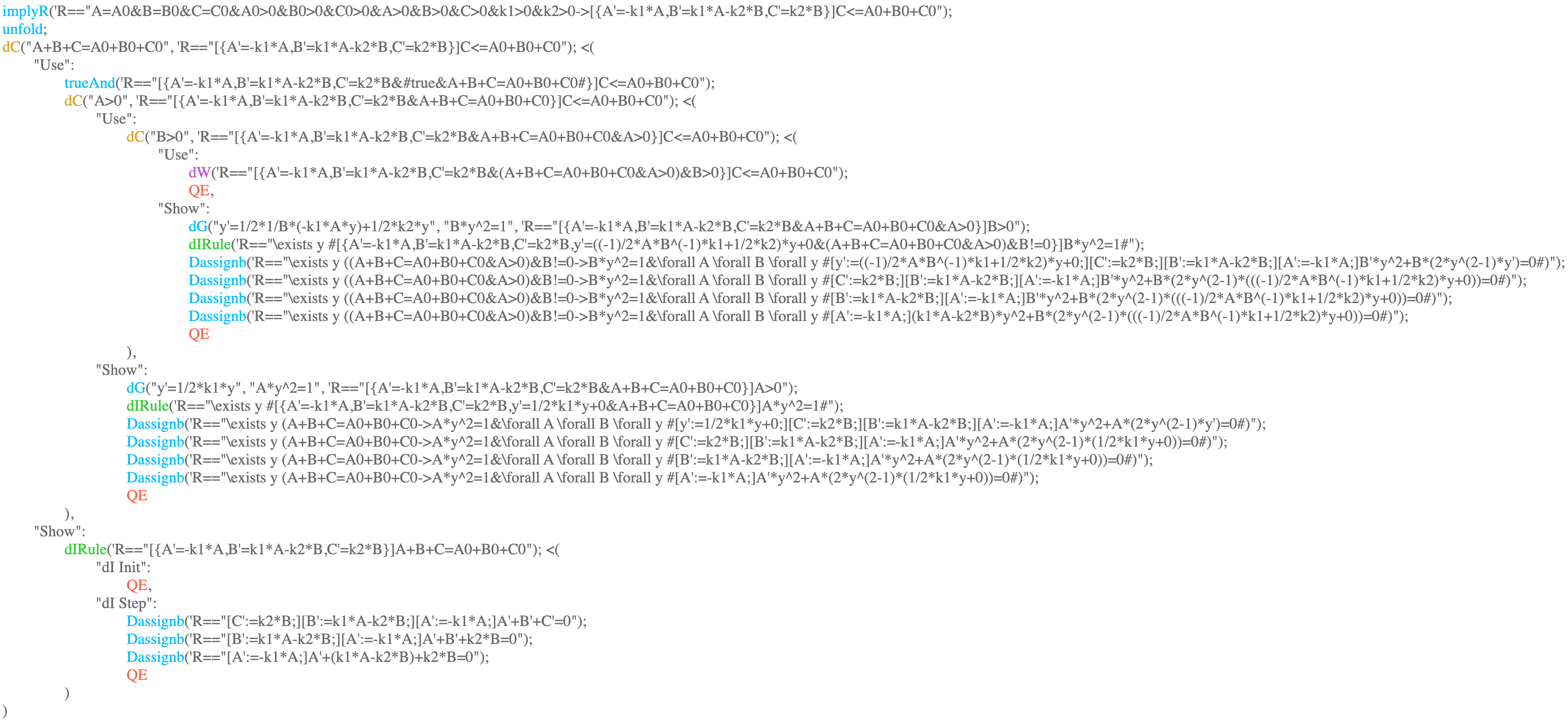} }}%
    \qquad
      \subfloat[\centering ]{{\includegraphics[width=11 cm]{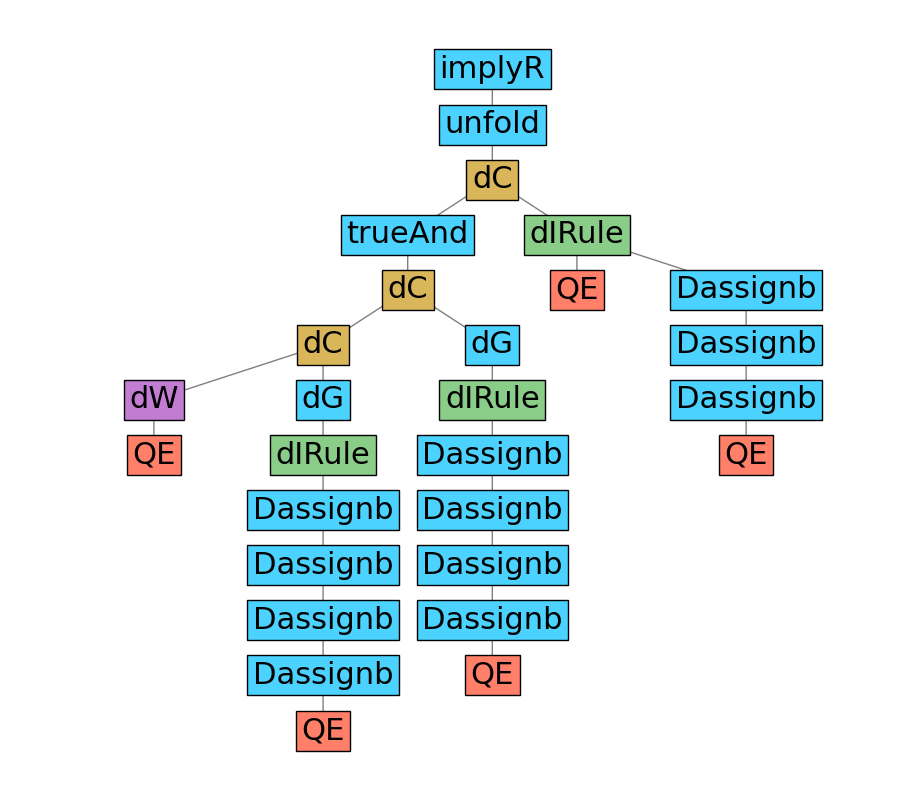} }}%
    \caption{a) full proof of two series reactions in batch reactor. b) Chart proof for two series reactions}
    \label{fig:Full-Chart-proof-series-reaction}%
\end{figure}

\end{document}